\documentclass{article}
\pdfoutput=1
\usepackage[margin=1.0in]{geometry}
\usepackage{graphicx}
\usepackage{bbm}
\usepackage{multicol}
\usepackage{subfigure}
\usepackage{float}
\usepackage{amsthm,amsmath,amssymb}
\usepackage{booktabs}
\usepackage{mathrsfs}
\usepackage{parskip}
\usepackage[]{hyperref}
\usepackage{makecell}
\usepackage{amsthm}
\usepackage{authblk}

\title{URIR: Recommendation algorithm of user RNN encoder and item encoder based on knowledge graph}
\author[1,2]{Na Zhao}
\author[1]{Zhen Long}
\author[3,4]{Zhi-Dan Zhao \thanks{Zhi-Dan Zhao: zzhidanzhao@gmail.com}}
\author[5]{Jian Wang \thanks{Jian Wang: jianwang@kust.edu.cn}}

\affil[1]{Key Laboratory in Software Engineering of Yunnan Province, School of Software, Yunnan University, Kunming 650504, People’s Republic of China}
\affil[2]{Electric Power Research Institute of Yunnan Power Grid Co., Ltd, Kunming 650200, People’s Republic of China}
\affil[3]{Department of Computer Science, School of Engineering, Shantou University, Shantou 515063, China}
\affil[4]{Key Laboratory of Intelligent Manufacturing Technology (Ministry of Education), Shantou University, Shantou 515063, China}
\affil[5]{College of Information Engineering and Automation, Kunming University of Science and Technology, Kunming 650217, People’s Republic of China}

\date{\today}
\newcommand{\tabcaption}{%
\setlength{\abovecaptionskip}{0pt}%
\setlength{\belowcaptionskip}{5pt}%
\caption}

\begin{document}

\maketitle

\begin{abstract}
Due to a large amount of information, it is difficult for users to find what they are interested in among the many choices. In order to improve users' experience, recommendation systems have been widely used in music recommendations, movie recommendations, online shopping, and other scenarios. Recently, Knowledge Graph (KG) has been proven to be an effective tool to improve the performance of recommendation systems. However, a huge challenge in applying knowledge graphs for recommendation is how to use knowledge graphs to obtain better user codes and item codes. In response to this problem, this research proposes a user Recurrent Neural Network (RNN) encoder and item encoder recommendation algorithm based on Knowledge Graph (URIR). This study encodes items by capturing high-level neighbor information to generate items' representation vectors and applies an RNN and items' representation vectors to encode users to generate users' representation vectors, and then perform inner product operation on users' representation vectors and items' representation vectors to get probabilities of users interaction with items. Numerical experiments on three real-world datasets demonstrate that URIR is superior performance to state-of-the-art algorithms in indicators such as AUC, Precision, Recall, and MRR. This implies that URIR can effectively use knowledge graph to obtain better user codes and item codes, thereby obtaining better recommendation results. 
    
\end{abstract}

\section{Introduction}
\label{intro}
With the rapid development of Internet, the amount of data has increased exponentially. Due to information overload, it is difficult for users to single out the ones they are interested in from a large number of choices~\cite{lmyzzz2012pr,guo2020Survey}. In order to make it easier for users to obtain the items and information they need, recommendation systems have been applied to scenarios such as music recommendation~\cite{hu2018Leveraging}, movie recommendation~\cite{wang2020setrank,zhuang2017representation}, and online shopping~\cite{han2019adaptive,xu2018exploiting}.
The recommendation algorithm is the core of recommendation system. One widely used recommendation algorithm is collaborative filtering (CF), which assigns representation vectors based on users' IDs and items' IDs~\cite{lmyzzz2012pr}. Then it models their interaction through specific operations (such as inner product~\cite{wang2017joint} or neural network~\cite{he2017neural}). In addition, methods based on CF are often plagued by the sparsity of users interaction with items and cold start problem. In order to solve these limitations, researchers usually turn to feature-rich scenarios, where the attributes of users and items are applied to compensate for the sparsity and improve performance of recommendation systems~\cite{cheng2016wide,wang2018shine,ZXZL:2021:IS,SLLC:2021:KBS}.

Some studies~\cite{huang2018improving,wang2018ripplenet,yu2014personalized,zhang2016collaborative,wang2019knowledge} go further than simply using attribute: they point out that attributes are not isolated but connected, which forms knowledge graph (KG). Generally, KG is a heterogeneous graph, where nodes correspond to entities (item or attribute of item), and edges correspond to relationships. Compared with KG-free methods, combining KG into recommendation has three benefits~\cite{wang2018ripplenet}: (1) The rich semantic relevance between items in KG helps to explore their potential connections and improve the accuracy of results; (2) The various types of relationships in a KG help to reasonably expand users' interests and increase the diversity of recommended items; (3) KG connects the items that users likes and the items recommended to users, which brings interpretability to recommendation systems.

The recommendation algorithm based on knowledge graph is mainly considered from the following three aspects: a) Method based on graph neural network. It captures the high-level structure in graph and refines the embedding of users and items. For example, RippleNet~\cite{wang2018ripplenet} spreads users' potential preferences in KG and explores their hierarchical interests. Wang et al.~\cite{wang2019knowledge} use a KG graph convolutional network (GCN~\cite{kipf2016semi}), which is incorporated to generate high-level term connectivity features. KGAT~\cite{wang2019kgat} explicitly models the high-level connections by an end-to-end manner. b) Method based on embedding, which combines the entities and relationships of KG into a continuous vector space, and helps recommendation system by enhancing semantic representation. For example, DKN~\cite{wang2018dkn} combines the semantic and knowledge-level representations of news, and incorporates KG representation into news recommendations. In addition, CKE~\cite{zhang2016collaborative} combines CF module with structure, text and visual knowledge into a unified recommendation framework. c) Method based on path. It uses paths and related user-item pairs to explore the connection mode between items in KG. For example, MCRec~\cite{hu2018Leveraging} learns the explicit representation of meta-paths in recommendation. In addition, it also considers meta-paths and the interactions between users and items. Compared with the method based on embedding, the method based on path develops  KG structure more naturally and intuitively. RKGE~\cite{sun2018recurrent} and KPRN~\cite{wang2019explainable} that needn't to design meta-paths automatically collect the paths between users and items, and use Recurrent Neural Network (RNN) to encode the paths.

However, the above algorithms have some problems such as the following: 1) users' representation vectors are obtained by simply coding in the methods based graph neural network and embedding, which may not be enough to encode users; 2) method based on embedding is more suitable for in-graph application, such as link prediction or KG completion, rather than recommendation algorithm; 3) method based on path  relies heavily on meta-paths and ignores the topological structure of the connections between users and items, and only aggregates the impact of each path on user preferences in the final stage. In order to solve the above-mentioned problems, this study propose a user RNN encoder and item encoder recommendation algorithm (URIR) based on knowledge graph. The algorithm aggregates and merges a fixed number of high-level neighborhood information with deviations when calculating the representation of a given entity in KG, and tries to encode users by an improved RNN and users' embeddings.

We apply URIR to three data sets: job (resume), ml (movie), and yelp (business). The experimental results show that URIR can obtain better recommendation results. Our contributions in this article are summarized as follows:\\
  a) We propose URIR, which is an end-to-end framework for exploring user preferences. By expanding the neighborhood of each entity in KG, URIR can capture the user's high-level personal interests.\\
  b) We conduct experiments on three recommendation scenarios, and the experimental results demonstrate the effectiveness of the URIR algorithm.\\
  c) We use RNN to encode users to obtain better users' representation vectors. As far as we know, it's the first time that RNN has been used to encode users in KG's algorithms.

The remainder of this paper is organized as follows: In section~\ref{matmeth}, we specify the Item Encoding Layer~\ref{item-encoding-layer}, the User Encoding Layer~\ref{user-encoding-layer}, the Prediction Layer~\ref{prediction-layer}, the Model Optimization~\ref{modelopt}, and the Experimental Setup~\ref{expset}. In section~\ref{result}, we conduct extensive experiments to evaluate the proposed method for URIR. Finally, in section~\ref{discuss}, we discuss and conclude this work.

\section{Material and methods}
\label{matmeth}
\begin{figure*}[!h]
\centering

\centering
\includegraphics[scale=0.4]{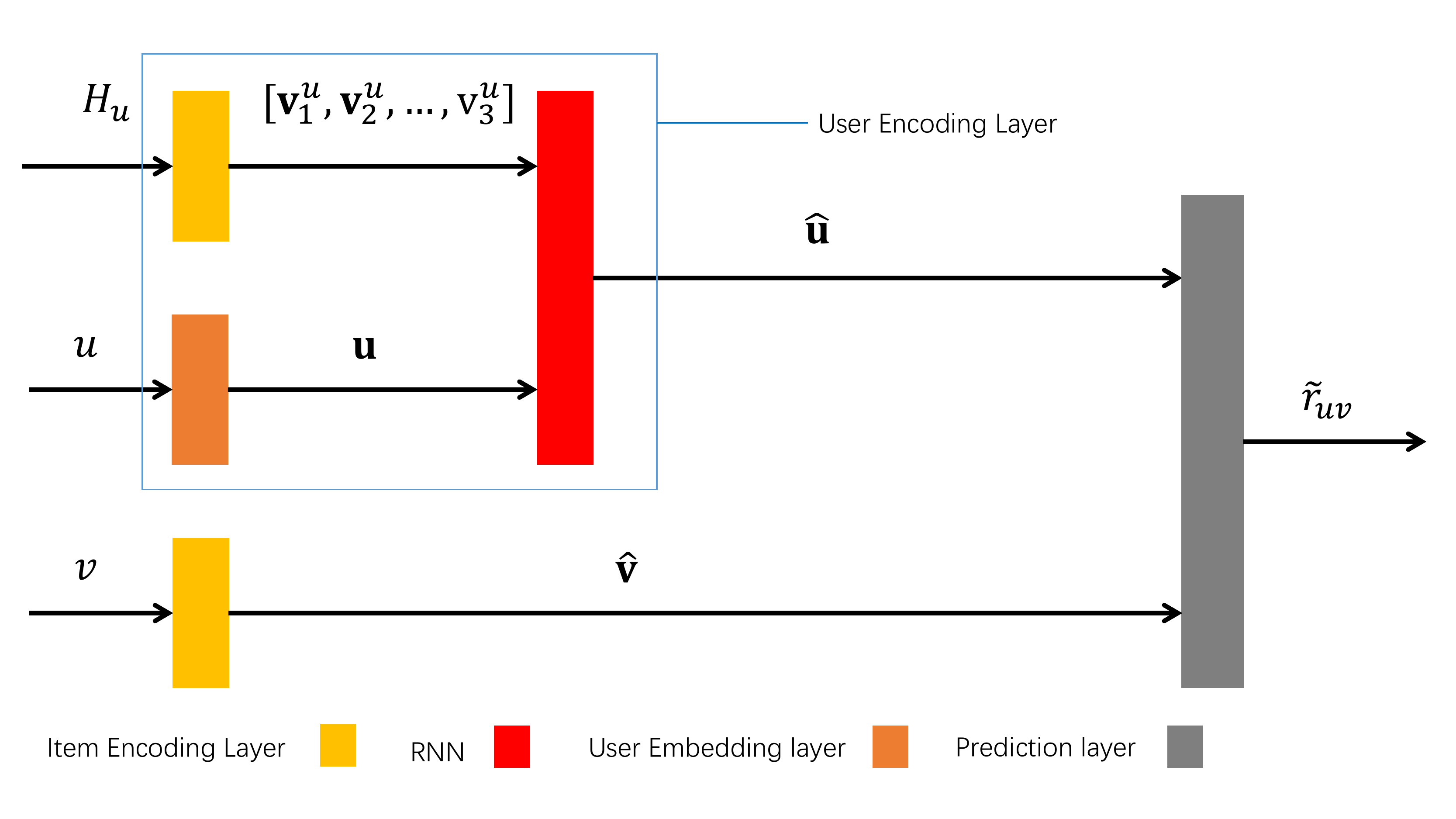}
\caption{Model Framework}
\label{fig:frameworkl}

\end{figure*}

We first introduce the commonly used mathematical symbols in the text and their meanings. The detailed information is shown in Table~\ref{tabsymbol}. Additionally,
the model framework is shown in Figure~\ref{fig:frameworkl}, which is mainly composed of 3 parts: 1) Item encoding layer~\ref{item-encoding-layer}, which responsible for encoding item $v$ and generating item $v$'s representation vector \textbf{\^{v}}. 2) User encoding layer~\ref{user-encoding-layer}, which illustrate the coding process of user $u$ and generate user $u$'s representation vector \textbf{\^{u}}. 3) Prediction layer~\ref{prediction-layer}, it generate the probability $\tilde{r}_{u,v}$ of the interaction between user $u$ and item $v$.


	\begin{table}[H]
	\caption{Model symbol introduction}
	\begin{center}
	\label{tabsymbol}
	\centering

    \begin{tabular}{|p{0.118\textwidth}<{\centering}|p{0.788\textwidth}<{\centering}|}%
    \hline
    \hline
    {\textbf{Notation}} & {\textbf{Description} } \\
	\hline
	$\mathcal{E}_v^l$ & \textit{v}'s l-level neighbsor\\
	${S}_v^l$ & \textit{v}'s l-level triples\\
	$(h,r,t)$ & KG's triples\\
	$\pi_{(h,r,t)}$ & $(h,r,t)$ weight\\
	$\tilde{\pi}_{(h,r,t)}$ & $(h,r,t)$ normalized weight\\
	$\mathbf{W}_*$ & learnable parameter  \\
	$\mathbf{b}_*$ & learnable parameter  \\
	$\mathbf{z}_{*}$ & hidden layer output \\
	$\mathbf{v}_l$ & $v$'s l-level neighbor vector \\
	$\hat{\mathbf{v}}$ & $v$'s representation vector\\
	$L$ & the order of the high-level neighbor \\
	$\mathbf{v}$ & $v$'s embedding \\
	$H_u$ & the set of the items that user $u$ has interacted with \\
	$r_{u,v}$ & {$r_{u,v}=1$ represents that user $u$ has interacted with item $v$, $r_{u,v}=0$ represents that user $u$ has not interacted with item $v$}\\
	$\hat{\mathbf{u}}$ & $u$'s representation vector \\
	$n$ & the number of elements of $H_u$ \\
	$\mathbf{v}_i^u$ & the representation of the item in $H_u$  \\
	$\mathbf{h}_i^u$ & $\mathbf{v}_i^u$'s RNN output  \\
    \hline
    \hline
    \end{tabular}
    \end{center}
	\end{table}

\subsection{Item Encoding Layer} \label{item-encoding-layer}
The purpose of item encoding layer is to capture the high-level information of item $v$ in knowledge graph and generate a better item's representation vector.
Here, item $v$ captures high-level neighbor information through \textit{v}'s l-level neighbors ($\mathcal{E}_v^l$) and \textit{v}'s l-level triples ($S_v^l$). As $l$ increases, the elements in $S_v^l$ increase exponentially. In order to reduce the calculation time, we collect $k$ neighbors of each node. This method is a commonly used simple but effective strategy to reduce computational cost~\cite{wang2018ripplenet,wang2019knowledge}.

	\begin{equation}
		\mathcal{E}_v^l=\{t|t\in\mathcal{G} \quad and  \quad (h,r,t)\in{S_v^l}\}\quad l=1,2,3,...
	\end{equation}
	\begin{equation}
		S_v^l=\{(h,r,t)|h\in\mathcal{E}_v^{l-1} \quad and  \quad (h,r,t)\in\mathcal{G}\quad and \quad t\notin\mathcal{E}_{v}^j\}\quad l=1,2,3,...\quad j<l
	\end{equation}
	\begin{equation}
		\mathcal{E}_v^0=\{v\}
	\end{equation}
	
We use triples in $S_v^l$ to capture high-level neighbor information. Due to the different head entities and relationships in KG, each tail entity of $S_v^l$ has different meanings and latent vector representations. For example, Forrest Gump and Cast Away have more in common in terms of their directors or casts, but are less similar if measured by genres or writers. As a result, we propose a knowledge-aware attentive embedding method to generate different attentive weights of the tail entity to reveal the different meanings, when it gets different head entities and relations.

	\begin{equation}
		\mathbf{z}_0=\mathbf{h \oplus r \oplus t} \label{equ:1}
	\end{equation}
	\begin{equation}
		\mathbf{z}_1={\rm{Relu}}(\mathbf{W}_1{\mathbf{z}_0}+\mathbf{b}_1)\label{equ:2}
	\end{equation}
	\begin{equation}
		\mathbf{z}_2={\rm{Relu}}(\mathbf{W}_2{\mathbf{z}_1}+\mathbf{b}_2)\label{equ:3}
	\end{equation}
	\begin{equation}
		\pi_{(h,r,t)}={\rm{Relu}}(\mathbf{W}_3{\mathbf{z}_2}+\mathbf{b}_3)\label{equ:4}
	\end{equation}
	\begin{equation}
		\tilde{\pi}_{(h,r,t)}=\frac{\exp(\pi_{(h,r,t)})}{\sum_{(h^{\prime},r^{\prime},t^{\prime})\in S_v^l}\exp(\pi_{(h^{\prime},r^{\prime},t^{\prime}))}}\label{equ:5}
	\end{equation}
	
where $\tilde{\pi}_{(h,r,t)}$ is the normalized weight that generated from $(h,r,t)$, ${\pi}_{(h,r,t)}$ is the weight that generated from $(h,r,t)$, $\mathbf h$ is the embedding of $h$, $\mathbf r$ is the embedding of $r$, $\mathbf t$ is the embedding of $t$, $\mathbf{W}_*$ is the learnable parameter, $\mathbf{b}_*$ is the learnable parameter, $\oplus$ is the concatenation of vector. 
	
Combine weights with corresponding tail entities to obtain high-level neighbor information, as shown in Eq.~\ref{equ:6}:
	\begin{equation}
		\mathbf{v}_l=\sum_{(h^{\prime},r^{\prime},t^{\prime})\in S_v^l}\tilde{\pi}_{(h^{\prime},r^{\prime},t^{\prime})}\mathbf{t}^{\prime}\label{equ:6}
	\end{equation}
	
where $\mathbf{v}_l$ is the $l$-level neighbor information of item $v$. Because the item itself has vital information, the high-level neighbor information and the information of the item itself are combined to obtain the item' representation vector, as shown in Eq.~\ref{equ:7}:
	\begin{equation}
		\hat{\mathbf{v}}=\frac{\sum_{i=1}^{L}\mathbf{v}_l+\mathbf{v}}{L+1}~\label{equ:7}
	\end{equation}
	
	where $\hat{\mathbf{v}}$ is the item $v$'s representation vector, $\mathbf v$ is the embedding of item $v$.
	
\subsection{User Encoding Layer} \label{user-encoding-layer}

The purpose of user encoding layer is to better encode users and generate better users' representation vectors. The user encoding layer consists of four parts: 1) item encoding layer: Input  the items that the user has interacted with, and then generate the corresponding item's representation vector; 2) user embedding layer: Generate users' embedding vectors; 4) RNN layer: Input  user's embedding vector and the representation vectors of the items in $H_u$ to obtain the user $u$'s representation vector. $H_u$ represent the set of the items that user $u$ has interacted with, as demonstrated in Eq.~\ref{equ:hu}.

	\begin{equation}
		H_u=\{v|r_{uv}=1\}
		\label{equ:hu}
	\end{equation}

Item encoding layer were introduced in~\ref{item-encoding-layer}, and user embedding layer is actually to obtain users' embedding. In order to improve the performance of the URIR, we use an improved RNN model. The improved RNN model is introduced as follows. URIR takes the representation vector of the item in $H_u$ as a word embedding vector and inputs them into the RNN. URIR considers the information carried by the user itself, that is, the user $u$'s embedding vector to encode user $u$. This is in line with daily habits. For example, two users have watched the same movie, but there is no guarantee that the movie they watch next time is the same. The RNN layer is represented by the following formulas for Eq.~\ref{equation:8} to~\ref{equation:10}:

	\begin{equation}
		\hat{\mathbf{u}}=\rm{RNN}([\mathbf{v}_1^{\emph{u}},\mathbf{v}_2^{\emph{u}},...,\mathbf{v}_\emph{n}^{\emph{u}}],\mathbf{u})\label{equation:8}
	\end{equation}
	\begin{equation}
		\mathbf{h}_i^u=\rm{ReLu}(\mathbf{W}\mathbf{v}_i^{\emph{u}}+\mathbf{H}\mathbf{h}_{i-1}^{\emph{u}}+\mathbf{U}\mathbf{u})\label{equation:9}
	\end{equation}
	\begin{equation}
		\hat{\mathbf{u}}=\mathbf{h}_n^{\emph{u}}\label{equation:10}
	\end{equation}
	
Where $\hat{\mathbf{u}}$ is the user $u$'s representation vector, $[\mathbf{v}_1^{\emph{u}},\mathbf{v}_2^{\emph{u}},...,\mathbf{v}_n^{\emph{u}}]$ are the representation vectors of the items in $H_u$, \textbf{u} is the user $u$'s embedding, $\mathbf{h}_0^u=\mathbf{0}$. URIR takes $\mathbf{h}_n^u$ as the user $u$'s representation vector, because it already contains the information previously entered.
	
\subsection{Prediction Layer} \label{prediction-layer}
Prediction layer obtains the probability of interaction between $u$ and $v$ through inner product of $\hat{\mathbf{u}}$ and $\hat{\mathbf{v}}$, and uses sigmoid function to control the probability to between 0 and 1, as shown in the following formula Eq.~\ref{equation:sigmoid}:
	\begin{equation}
		\tilde{r}_{u,v}=\rm{sigmoid}(\hat{\mathbf{u}}\hat{\mathbf{v}}^{\rm{T}})\label{equation:sigmoid}
	\end{equation}
	
	Where $\tilde{r}_{u,v}$ is the predicted probability of interaction between user $u$ and item $v$.
	
\subsection{Model Optimization}\label{modelopt}
Objective function: According to the literature~\cite{he2017neural}, we treat top-K recommendation task as a binary classification problem, where 1 means that user has interacted with item, and 0 means that user has not interacted with item. In objective function, we use cross entropy as objective function, the formula is as follows Eq.~\ref{equation:modfunc}:
	\begin{equation}
		\mathcal{J}=\sum_{(u,v)\in\mathcal{D}_{train}}\rm{BCELOSS}(\tilde{\emph{r}}_\emph{u,v},\emph{r}_\emph{u,v})+\lambda||\mathcal{F}||_2^2\label{equation:modfunc}
	\end{equation}
	
Where $\lambda$ is L2 regularization coefficient, $\mathcal{F}$ are the parameters of the model, and $\mathcal{D}_{train}$ is training set. We collect the items that have not been interacted with for each user as negative samples, and negative samples are collected according to the ratio of the number of positive samples to the number of negative samples of 4:1 in training set.
	
\subsection{Experimental Setup}\label{expset}
To verify the performance of algorithm URIR, We use three data sets: (1) Job data set describes some occupations that users have carried on; (2) Ml~\cite{sun2018recurrent} is a data set widely used in movie recommendation; (3) Yelp~\cite{sun2018recurrent} records users ratings of some businesses. In order to better verify the performance of the model in solving the cold start problem, we limit the number of user interactions with items to less than 20~\cite{wang2019kgat}. The attributes of the data sets are shown in Table~\ref{table_1}. Users represents the number of users, items represents the number of items, relations represents the number of relation types, edges represents the number of edges, interactions represents the number of interactions, entities represents the number of entities (nodes).
	\begin{table}[H]
	\tabcaption{Data set attributes}
	
	\label{table_1}
	\centering
	\begin{tabular}{ccccccc}
	\toprule
	data set&users&items&relations&edges&interactions&entities  \\
	\midrule
	job&5967&85&2&360&13220&194   \\
	ml&943&1674&3&13202&12434&6800 \\
	yelp&1936&11515&2&42875&18146&12097\\
	\bottomrule
	\end{tabular}
	\end{table}
	
Evaluation indicators: We use widely applied indicators to evaluate models' performance. For each user, we divide him\verb|\|her history records according to the ratio of the number of items in the training set to the number of items in the test set of 7:3, and We collect negative samples according to the ratio of the number of positive samples to the number of negative samples of 4:1 in training set.
	
According to the literature~\cite{he2017neural,wang2019explainable,elkahky2015multi}, in the test process, for each user, we randomly select 50 items that the user has not interacted with, and randomly select 1 item that has interacted with the user in test set. The recommended list is composed of these 51 items, and the 51 items are sorted to reduce test's complexity. In model training, we use AUC~\cite{evaluation} as an evaluation indicator to determine the best parameters. When evaluating model performance, we use Precision@K~\cite{evaluation}, Recall@K~\cite{evaluation}and MRR@K~\cite{evaluation} as evaluation indicators. We calculate these indicators for each user, and calculate the average value when K = \{1, 2, 4, 5, 6, 8, 10\}. Since situation does not recommend too many items in actual, we choose these values for K. Generally, higher values of these indicators indicate better performance. AUC characterizes the overall performance of recommendation algorithm and covers the performance of all different recommendation list lengths~\cite{evaluation}. Therefore, we use AUC as an evaluation indicator during training.

In order to verify the effectiveness of URIR, we select state-of-the-art algorithms recently proposed for comparison. These algorithms are: NFM~\cite{he2017neural2}, KPRN~\cite{wang2019explainable}, Wide\&Deep~\cite{cheng2016wide}, RKGE~\cite{sun2018recurrent}, RippleNet~\cite{wang2018ripplenet}, KGCN~\cite{wang2019knowledge}. These algorithms were chosen because they are all influential algorithms based on knowledge graph, and some of the algorithm ideas have provided useful inspiration for us to propose the URIR algorithm.

Parameter settings: The optimal parameters of the above algorithms are set by experimental verification or the conclusions obtained of the original paper~\cite{he2017neural2,wang2019explainable,cheng2016wide,sun2018recurrent,wang2018ripplenet,wang2019knowledge}. For URIR, we use Adam as optimizer, and use mini-batch gradient descent algorithm. These methods are some commonly used optimization strategies~\cite{wang2018ripplenet,wang2019kgat,wang2019knowledge}. In view of the influence of different hyperparameters the embedding dimension of entity \textit{d}, the number of neighbors collected \textit{k}, the order of the high-level neighbor \textit{L} and the number of elements of $H_u$ \textit{n}, we also made a comprehensive comparison of the system. The learning rate, batch size, and L2 regularization coefficient are determined according to training time. The initialization parameters of the model are: \textit{d}=4, \textit{k}=4, \textit{L}=2, \textit{n}=5. Hyperparameter adjustment sequence is \textit{d}, \textit{L}, \textit{k}, \textit{n}. The detailed comparison process and comparison results are introduced in Section~\ref{parameter-sensitivity}. The default values of these parameters are shown in Table~\ref{tab:parameter}.

	\begin{table}[H]
	\tabcaption{Parameter settings on each data set}
	
	\label{tab:parameter}
	\centering
	\begin{tabular}{ccccccccc}
	\toprule
	data set&learning rate&number of iterations&\textit{d}&\textit{k}&\textit{L}&\textit{n}&$\lambda$&batch size  \\
	\midrule
	job & 0.02 & 11 & 8  & 4  & 1 & 5 & 0.001 &256  \\
	ml  & 0.05 & 10 & 64 & 16 & 1 & 5 & 0.0001&256 \\
	yelp& 0.05 & 6  & 64 & 16 & 1 & 5 & 0.0001&256     \\
	\bottomrule
	\end{tabular}
	\end{table}
	
\section{Results}\label{result}

We conducted extensive experiments on three data sets to answer the following three questions: 1) Does the URIR model perform better than other advanced recommendation algorithms? 2) What is the impact of RNN on the performance of the URIR algorithm? 3) How does the hyperparameter affect the performance of the URIR?

\subsection{Performance of URIR}
\begin{figure}[h]
	
	\centering
    \includegraphics[scale=0.418]{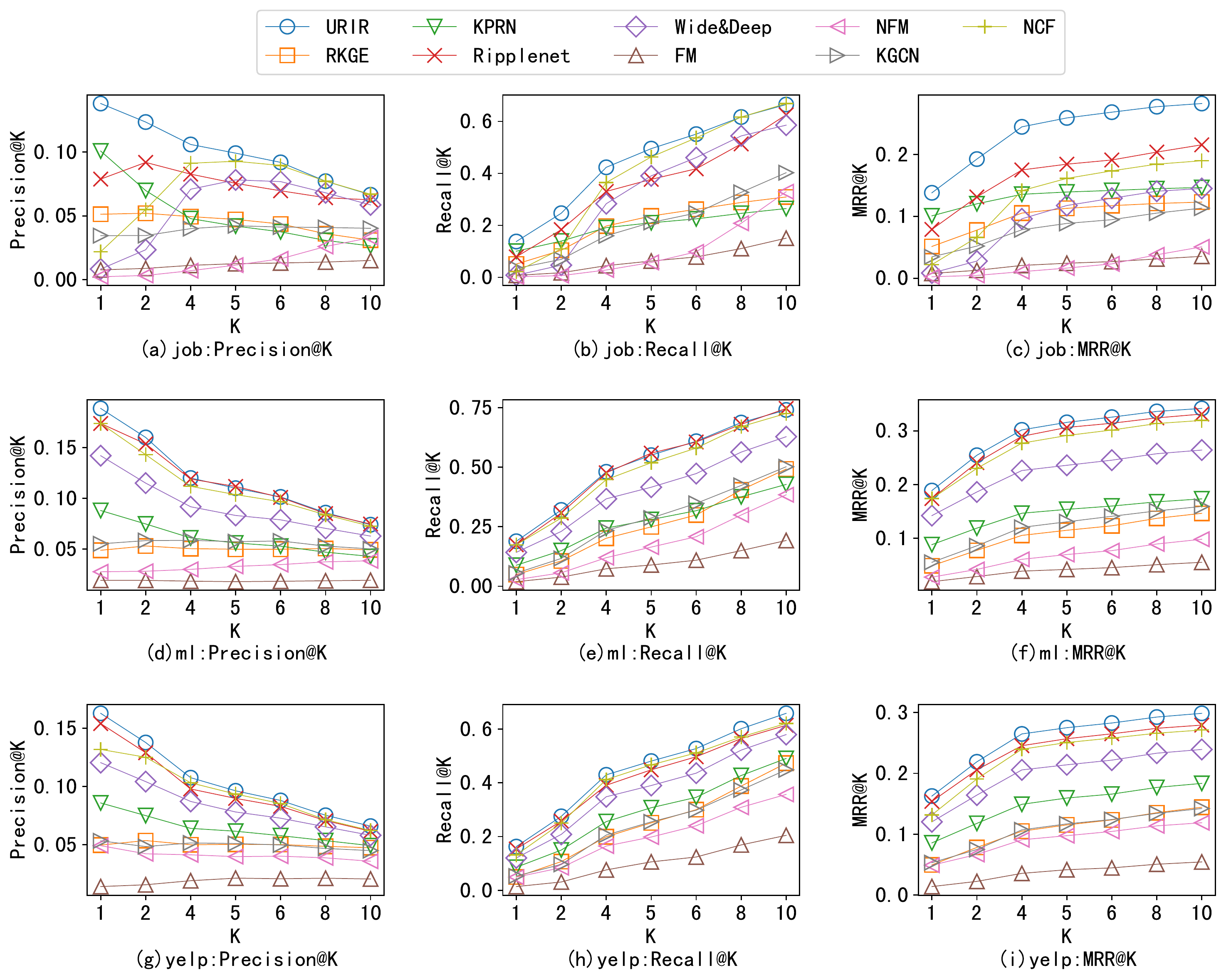}
	    \caption{URIR's performance on three data sets}
	    \label{fig:comparsion}
	\end{figure}

In order to fully understand the performance of the URIR algorithm, we compared the performance of the aforementioned eight different algorithms on three different data sets, and the results are shown in Figure~\ref{fig:comparsion}. First of all, Figure~\ref{fig:comparsion}~(a), Figure~\ref{fig:comparsion}~(d) and Figure~\ref{fig:comparsion}~(g) display the variation trend of the Precision of different algorithms on three different data sets as K increases, respectively. All the algorithm result curves in Figure~\ref{fig:comparsion}~(a) show that as K increases, the overall Precision shows a downward trend. Although the URIR algorithm curve and other algorithm curves is getting closer, the performance of URIR is still better than other algorithms on the whole. For example, when K=\{1, 2, 4, 5, 6, 8, 10\}, compared with the best results of state-of-the-art algorithms, URIR increases by 26.9\%, 36.9\%, 5.5\%, 1.7\% and 0.0\%, respectively. Additionally, compared with the worst case of these algorithms, when K=\{1, 2, 4, 5, 6, 8, 10\}, URIR increased by 147.5\%, 190\%, 28.9\%, 9.6\% and 0.9\%, respectively. Compared with the average values of other models when K=\{1, 2, 4, 5, 6, 8, 10\}, URIR increased by 61.0\%, 64.9\%, 14.6\%, 4.3\% and 0.5\%, respectively. Similarly, both the ml and yelp data sets show the same characteristics. This phenomenon is consistent with previous studies~\cite{wang2018ripplenet,wang2019kgat,wang2019knowledge,WGTQ:2020:CKAN}.

Secondly, Figure~\ref{fig:comparsion}~(b), Figure~\ref{fig:comparsion}~(e) and Figure~\ref{fig:comparsion}~(h) demonstrate the evolution trend of Recall and K of different algorithms on 3 different data sets. For example, Figure~\ref{fig:comparsion}~(b) shows that the Recall of different algorithms on the job data set shows an overall upward trend as K increases. Although the Recall of all algorithms shows a trend of increasing with K, the Recall value of the URIR algorithm is always higher than that of other methods.

Thirdly,  Figure~\ref{fig:comparsion}~(c), Figure~\ref{fig:comparsion}~(f) and Figure~\ref{fig:comparsion}~(i) illustrate the changes in the MRR of different algorithms on three different data sets as K increases. Figure~\ref{fig:comparsion}~(c) shows the changing trend of MRR of different algorithms on the job data set. It is very interesting that we observe that all the curves in Figure~\ref{fig:comparsion}~(c) show an upward trend as K increases. At the same time, although all models have the same trend, the MRR value of the URIR algorithm has always been higher than other models. For example, when URIR is K=\{1, 2, 4, 5, 6, 8, 10\}, compared with the best model, it is improved by 26.9\%, 43.1\%, 32.8\%, 32.1\% and 31.8\%, respectively. Compared with the worst model when K=\{1, 2, 4, 5, 6, 8, 10\}, URIR increased by 147.5\%, 168.4\%, 114.1\%, 100.8\% and 100.0\%, respectively. Compared with the average value of other models, When K=\{1, 2, 4, 5, 6, 8, 10\}, URIR increased by 61.0\%, 74.1\%, 56.9\%, 53.9\% and 53.0\%, respectively.

In general, our experimental results show that compared with other models, URIR can better find items that users are interested in and put them at the top of the recommendation list. For example, if the user only requests to recommend fewer items, compared to other algorithms, URIR better meets the needs of users. In addition, as the value of K increases, the advantage of the URIR algorithm gradually becomes smaller. This is due to URIR put the items that users like at the top of the recommended list. As K increases, other algorithms also predict the items users like. In addition, when using ml data, the performance of URIR is not as good as the other two data. The possible factor is that there are enough interaction records between users and items of ml data. There is no need to use RNN to capture more information to encode users, which implies that the URIR algorithm is a useful attempt to solve the cold start problem of collaborative filtering recommendation algorithm.

\subsection{The impact of RNN}\label{rnnimpact}
In order to generate a better user representation vector, the URIR model considers the introduction of RNN to encode users. Therefore, whether the use of RNN has an impact on user coding is an important issue. Here, we answer the above question by comparing the recommendation results of URIR and the URIR algorithm without the RNN (URIR-RNN). At this time, user $u$'s representation vector is $\hat{\mathbf{u}}=\mathbf{u}+\sum_{v \in H_u}\hat{\mathbf{v}}$. The experimental results are shown in Table \ref{tab:URIR-RNN}. The experimental results are shown in Table~\ref{tab:URIR-RNN}. The research results indicate that after the application of RNN user encoding, the AUC results of the two datasets job and ml of URIR have been significantly improved. Especially on the job data set, the AUC value of the URIR algorithm is about 0.84, and the AUC value of the URIR-RNN is about 0.75, an increase of up to 10\%. This result verifies that the use of RNN proposed in URIR to encode users is very necessary and effective. On Yelp data, the reason why the AUC values of the two algorithms are not significantly different may be that there are many interaction records between users and items in the yelp data and the KG of yelp is so big that URIR-RNN can get enough auxiliary information from the KG. The result also reflects that the item encoder of the URIR algorithm effectively obtain auxiliary information from KG.

	\begin{table*}[h]
		\tabcaption{Comparison of AUC between URIR and URIR-RNN}
		\label{tab:URIR-RNN}
		\centering
		\begin{tabular}{cccc}
		\toprule
			model    & job    & ml     & yelp \\
			\midrule
			URIR     & 0.8351                 & 0.8544                 & 0.8039 \\
			URIR-RNN & 0.7544(\textbf{-9.66\%}) & 0.8059(\textbf{-5.68\%}) & 0.8072 (\textbf{0.41\%})\\
			\bottomrule
		\end{tabular}
	\end{table*}

\subsection{Parameter Sensitivity}\label{parameter-sensitivity}
In order to better understand the role of different hyperparameters in the URIR algorithm, here we separately study the impact of these parameter changes on the performance of the URIR algorithm: the embedding dimension of entity $d$, the order of the high-level neighbor \textit{L}, the number of neighbors collected \textit{k} and the number of elements of $H_u$ $n$.

	\begin{table*}[!h]

	\tabcaption{Comparison of AUC for different $d$ of URIR}
	
	\label{tab:d}
	\centering
	\begin{tabular}{cccccc}
	\toprule
	data set & 4      & 8 & 16 & 32 & 64 \\
	\midrule
	job      & 0.8031 & {\textbf{0.8100}} & 0.7879 & 0.7588 & 0.8039 \\
	ml       & 0.8160 & 0.8132 & 0.8248 & 0.8257 & {\textbf {0.8292}} \\
	yelp     & 0.7761 & 0.7751 & 0.7805 & 0.7900 & {\textbf {0.7967}} \\
	\bottomrule
	\end{tabular}
	\end{table*}

Embedding dimension of entity $d$: The role of entity embedding dimension $d$ in the knowledge graph is a general problem~\cite{wang2018ripplenet,wang2019kgat,wang2019knowledge,WGTQ:2020:CKAN}. Based on this assumption, we give the performance of $d$ taking different values \{4, 8, 16, 32, 64\} in the URIR algorithm, and the experimental results are shown in table~\ref{tab:d}. We noticed that the performance of URIR increase with the $d$ growing, and reach the best performance when $d$ reaches a certain value. This result implies that a larger embedding dimension $d$ can significantly help to encode more useful information. However, with the further increase of $d$, the performance of some data sets decrease, such as job data sets. As shown in Table~\ref{table_1}, the sparsity of the job graph is significantly smaller than the other two data of ml and yelp. Future research should focus on the influence of graph sparsity on the selection of embedding dimensions~\cite{wang2019knowledge,WGTQ:2020:CKAN}. This result indicates that too large embedding dimensions may cause the entity to be overrepresent and introduce noise, which affects the performance of the algorithm.

	\begin{table*}[h]

	\tabcaption{Comparison of AUC for different $L$ of URIR}
	
	\label{tab:L}
	\centering
	\begin{tabular}{ccccc}
	\toprule
	data set & 1      & 2 & 3 & 4 \\
	\midrule
	job      & {\textbf{0.8351}} & 0.8100 & 0.8049 & 0.7938 \\
	ml       & {\textbf{0.8542}} & 0.8292 & 0.8377 & 0.8375  \\
	yelp     & {\textbf{0.7991}} & 0.7967 & 0.7830 & 0.7789  \\
	\bottomrule
	\end{tabular}
	\end{table*}
	
The order of the high-level neighbor \textit{L}: We studied the impact of the order of the high-level neighbor \textit{L} on URIR. The results are shown in Table~\ref{tab:L}. The experimental results show that when $L=1$, the AUC value reaches the highest value in all three data sets. This phenomenon is in line with our intuition. Because as $L$ gets larger, the less node information the project $v$ can obtain. However, the correlation between further nodes and $v$ will not be very high, but some noise is transmitted to item $v$'s representation vector. Another reason is the role of the RNN encoder. Because URIR-RNN's $L$ is not always one.
	
	\begin{table}[H]

	\tabcaption{Comparison of AUC for different $k$ of URIR}
	
	\label{tab:k}
	\centering
	\begin{tabular}{ccccc}
	\toprule
	data set & 2      & 4 & 8 & 16 \\
	\midrule
	job      & 0.8231 & {\textbf{0.8351}} & 0.8181 & 0.0.8193 \\
	ml       & 0.8429 & 0.8542 & 0.8523 & {\textbf{0.8544}}  \\
	yelp     & 0.7755 & 0.7991 & 0.8032 & {\textbf{0.8039}}  \\
	\bottomrule
	\end{tabular}
	\end{table}

The number of neighbors collected \textit{k}: The influence of the number of neighbors on the effect of the algorithm is a hot research issue~\cite{wang2018ripplenet,wang2019knowledge,WGTQ:2020:CKAN}. In this section, we study the influence of the number of neighbors on the recommendation based on the KG by changing the number of neighbors collected \textit{k}. The results are shown in the table~\ref{tab:k}. The experimental results clearly show that as the number of neighbors k increases, the AUC value also increases. This is because the small $k$ does not have enough capacity to merge neighborhood information. However, it is observed that the AUC value in the job data set becomes smaller. This may be similar to the reason for the dimension $d$, that is, the job data set is too sparse, and increasing the value of k cannot obtain enough domain information, but as k increases, it brings noise .

	\begin{table}[H]

	\tabcaption{Comparison of AUC for different $n$ of URIR}
	
	\label{tab:n}
	\centering
	\begin{tabular}{ccccc}
	\toprule
	data set & 5      & 10 & 15 & 20 \\
	\midrule
	job      & {\textbf{0.8351}} & 0.8341 & 0.8341 & 0.8341 \\
	ml       & {\textbf{0.8544}} & 0.8488 & 0.8472 & 0.8472  \\
	yelp     & {\textbf{0.8039}} & 0.8037 & 0.8038 & 0.8038  \\
	\bottomrule
	\end{tabular}
	\end{table}
	
The number of elements of $H_u$ $n$: We studied the impact of the number of elements of $H_u$ $n$. The results are shown in Table \ref{tab:n}. When $n$=5, AUC is the highest. This is not difficult to understand. Because we speculate people's preferences from a few things. For example, if a user has watched four movies that a actor stars, we can infer that the user likes the actor. However, the user not only watched these movies, but also watched a certain director's movie. The actor also star the movie. We may also think that the user like the director and recommend the director's movie. 
However, the user only likes the actor and does not like the director in actual. In other words, an appropriate $n$ can help the model understand the user's preferences. But an excessively large $n$ makes the model misunderstand the user's preferences.

\section{Discussion and conclusion}
\label{discuss}

This paper proposes a recommendation algorithm for user RNN encoder and item encoder based on knowledge graph, which integrates the structured information of knowledge graph into the algorithm. The RNN encoder helps to understand users' preferences, and the item encoder helps to mine $v$'s high-level information. These are beneficial to the generation of better users' representation vectors and items' representation vectors, make full use of knowledge graphs, and solve the limitations of the three methods. URIR effectively alleviates the sparsity problem and improves the accuracy of recommendation systems. Experiments were carried out on 3 real-world data sets. we use AUC, Precision, Recall and MRR as the evaluation indicators of the models. URIR was compared with FM, NFM~\cite{he2017neural2}, KPRN~\cite{wang2019explainable}, Wide\&Deep~\cite{cheng2016wide}, RKGE~\cite{sun2018recurrent}, RippleNet~\cite{wang2018ripplenet}, KGCN~\cite{wang2019knowledge}. Our experiment results demonstrate that URIR is generally better than the advanced algorithm on three data sets, can provide users with more accurate recommendation results, and has strong flexibility in fusing heterogeneous data. Good performance in AUC, Precision and Recall shows that it can be applied to classification tasks or tasks where users rate items. Achieving better performance in MRR indicates that it can be applied to the task of sorting items.

The URIR algorithm understands users' preferences through users' historical behaviors, which is very dependent on whether the historical behaviors collected contain all preferences and contains redundant information. Future work should focus on how to improve the sampling strategy and user encoding layer to generate better users' representation vectors. In addition, future recommendation methods based on knowledge graphs should include specific network details, including user rating deviations and rating features~\cite{SLLC:2021:KBS}, network node degrees and H-index~\cite{ZLTC:2018:EPL}, and hypernetwork loop structure~\cite{PSLDW:2021:EPL} etc.

	\bibliography{URIR}

\begin{thebibliography}{10}

\bibitem{lmyzzz2012pr}
Linyuan Lü, Matúš Medo, Chi~Ho Yeung, Yi-Cheng Zhang, Zi-Ke Zhang, and Tao
  Zhou.
\newblock Recommender systems.
\newblock {\em Physics Reports}, 519(1):1--49, 2012.

\bibitem{guo2020Survey}
Qingyu Guo, Fuzhen Zhuang, Chuan Qin, Hengshu Zhu, Xing Xie, Hui Xiong, and
  Qing He.
\newblock A survey on knowledge graph-based recommender systems.
\newblock {\em IEEE Transactions on Knowledge and Data Engineering}, 2020.

\bibitem{hu2018Leveraging}
Binbin Hu, Chuan Shi, Wayne~Xin Zhao, and Philip~S. Yu.
\newblock Leveraging meta-path based context for top- n recommendation with a
  neural co-attention model.
\newblock In {\em Proceedings of the 24th ACM SIGKDD International Conference
  on Knowledge Discovery \&amp; Data Mining}, KDD '18, page 1531–1540, New
  York, NY, USA, 2018. Association for Computing Machinery.

\bibitem{wang2020setrank}
Chao Wang, Hengshu Zhu, Chen Zhu, Chuan Qin, and Hui Xiong.
\newblock Setrank: A setwise bayesian approach for collaborative ranking from
  implicit feedback.
\newblock In {\em Proceedings of the AAAI Conference on Artificial
  Intelligence}, number~04, pages 6127--6136, 2020.

\bibitem{zhuang2017representation}
Fuzhen Zhuang, Zhiqiang Zhang, Mingda Qian, Chuan Shi, Xing Xie, and Qing He.
\newblock Representation learning via dual-autoencoder for recommendation.
\newblock {\em Neural Networks}, 90:83--89, 2017.

\bibitem{han2019adaptive}
Jiayu Han, Lei Zheng, Yuanbo Xu, Bangzuo Zhang, Fuzhen Zhuang, S~Yu Philip, and
  Wanli Zuo.
\newblock Adaptive deep modeling of users and items using side information for
  recommendation.
\newblock {\em IEEE transactions on neural networks and learning systems},
  31(3):737--748, 2019.

\bibitem{xu2018exploiting}
Yuanbo Xu, Yongjian Yang, Jiayu Han, En~Wang, Fuzhen Zhuang, and Hui Xiong.
\newblock Exploiting the sentimental bias between ratings and reviews for
  enhancing recommendation.
\newblock In {\em 2018 IEEE International Conference on Data Mining (ICDM)},
  pages 1356--1361. IEEE, 2018.

\bibitem{wang2017joint}
Hongwei Wang, Jia Wang, Miao Zhao, Jiannong Cao, and Minyi Guo.
\newblock Joint topic-semantic-aware social recommendation for online voting.
\newblock In {\em Proceedings of the 2017 ACM on Conference on Information and
  Knowledge Management}, pages 347--356, 2017.

\bibitem{he2017neural}
Xiangnan He, Lizi Liao, Hanwang Zhang, Liqiang Nie, Xia Hu, and Tat-Seng Chua.
\newblock Neural collaborative filtering.
\newblock In {\em Proceedings of the 26th international conference on world
  wide web}, pages 173--182, 2017.

\bibitem{cheng2016wide}
Heng-Tze Cheng, Levent Koc, Jeremiah Harmsen, Tal Shaked, Tushar Chandra,
  Hrishi Aradhye, Glen Anderson, Greg Corrado, Wei Chai, Mustafa Ispir, et~al.
\newblock Wide \& deep learning for recommender systems.
\newblock In {\em Proceedings of the 1st workshop on deep learning for
  recommender systems}, pages 7--10, 2016.

\bibitem{wang2018shine}
Hongwei Wang, Fuzheng Zhang, Min Hou, Xing Xie, Minyi Guo, and Qi~Liu.
\newblock Shine: Signed heterogeneous information network embedding for
  sentiment link prediction.
\newblock In {\em Proceedings of the Eleventh ACM International Conference on
  Web Search and Data Mining}, pages 592--600, 2018.

\bibitem{ZXZL:2021:IS}
Ming-yang Zhou, Rong-qin Xu, Zi-ming Wang, and Hao Liao.
\newblock A generic bayesian-based framework for enhancing top-n recommender
  algorithms.
\newblock {\em Information Sciences}, 580:460--477, 2021.

\bibitem{SLLC:2021:KBS}
Hong-Liang Sun, Kai-Ping Liang, Hao Liao, and Duan-Bing Chen.
\newblock Evaluating user reputation of online rating systems by rating
  statistical patterns.
\newblock {\em Knowledge-Based Systems}, 219:106895, 2021.

\bibitem{huang2018improving}
Jin Huang, Wayne~Xin Zhao, Hongjian Dou, Ji-Rong Wen, and Edward~Y Chang.
\newblock Improving sequential recommendation with knowledge-enhanced memory
  networks.
\newblock In {\em The 41st International ACM SIGIR Conference on Research \&
  Development in Information Retrieval}, pages 505--514, 2018.

\bibitem{wang2018ripplenet}
Hongwei Wang, Fuzheng Zhang, Jialin Wang, Miao Zhao, Wenjie Li, Xing Xie, and
  Minyi Guo.
\newblock Ripplenet: Propagating user preferences on the knowledge graph for
  recommender systems.
\newblock In {\em Proceedings of the 27th ACM International Conference on
  Information and Knowledge Management}, pages 417--426, 2018.

\bibitem{yu2014personalized}
Xiao Yu, Xiang Ren, Yizhou Sun, Quanquan Gu, Bradley Sturt, Urvashi Khandelwal,
  Brandon Norick, and Jiawei Han.
\newblock Personalized entity recommendation: A heterogeneous information
  network approach.
\newblock In {\em Proceedings of the 7th ACM international conference on Web
  search and data mining}, pages 283--292, 2014.

\bibitem{zhang2016collaborative}
Fuzheng Zhang, Nicholas~Jing Yuan, Defu Lian, Xing Xie, and Wei-Ying Ma.
\newblock Collaborative knowledge base embedding for recommender systems.
\newblock In {\em Proceedings of the 22nd ACM SIGKDD international conference
  on knowledge discovery and data mining}, pages 353--362, 2016.

\bibitem{wang2019knowledge}
Hongwei Wang, Miao Zhao, Xing Xie, Wenjie Li, and Minyi Guo.
\newblock Knowledge graph convolutional networks for recommender systems. corr
  abs/1904.12575 (2019).
\newblock {\em arXiv preprint arXiv:1904.12575}, 2019.

\bibitem{kipf2016semi}
Thomas~N Kipf and Max Welling.
\newblock Semi-supervised classification with graph convolutional networks.
\newblock {\em arXiv preprint arXiv:1609.02907}, 2016.

\bibitem{wang2019kgat}
Xiang Wang, Xiangnan He, Yixin Cao, Meng Liu, and Tat-Seng Chua.
\newblock Kgat: Knowledge graph attention network for recommendation.
\newblock In {\em Proceedings of the 25th ACM SIGKDD International Conference
  on Knowledge Discovery \& Data Mining}, pages 950--958, 2019.

\bibitem{wang2018dkn}
Hongwei Wang, Fuzheng Zhang, Xing Xie, and Minyi Guo.
\newblock Dkn: Deep knowledge-aware network for news recommendation.
\newblock In {\em Proceedings of the 2018 world wide web conference}, pages
  1835--1844, 2018.

\bibitem{sun2018recurrent}
Zhu Sun, Jie Yang, Jie Zhang, Alessandro Bozzon, Long-Kai Huang, and Chi Xu.
\newblock Recurrent knowledge graph embedding for effective recommendation.
\newblock In {\em Proceedings of the 12th ACM Conference on Recommender
  Systems}, pages 297--305, 2018.

\bibitem{wang2019explainable}
Xiang Wang, Dingxian Wang, Canran Xu, Xiangnan He, Yixin Cao, and Tat-Seng
  Chua.
\newblock Explainable reasoning over knowledge graphs for recommendation.
\newblock In {\em Proceedings of the AAAI Conference on Artificial
  Intelligence}, volume~33, pages 5329--5336, 2019.

\bibitem{elkahky2015multi}
Ali~Mamdouh Elkahky, Yang Song, and Xiaodong He.
\newblock A multi-view deep learning approach for cross domain user modeling in
  recommendation systems.
\newblock In {\em Proceedings of the 24th international conference on world
  wide web}, pages 278--288, 2015.

\bibitem{evaluation}
Yuxiao Zhu and Linyuan Lü.
\newblock Overview of recommendation system evaluation indicators.
\newblock {\em Journal of University of Electronic Science and Technology of
  China}, 02:163--175, 2012.

\bibitem{he2017neural2}
Xiangnan He and Tat-Seng Chua.
\newblock Neural factorization machines for sparse predictive analytics.
\newblock In {\em Proceedings of the 40th International ACM SIGIR conference on
  Research and Development in Information Retrieval}, pages 355--364, 2017.

\bibitem{WGTQ:2020:CKAN}
Ze~Wang, Guangyan Lin, Huobin Tan, Qinghong Chen, and Xiyang Liu.
\newblock Ckan: Collaborative knowledge-aware attentive network for recommender
  systems.
\newblock In {\em Proceedings of the 43rd International ACM SIGIR Conference on
  Research and Development in Information Retrieval}, SIGIR '20, page
  219–228, New York, NY, USA, 2020. Association for Computing Machinery.

\bibitem{ZLTC:2018:EPL}
Xuzhen Zhu, Wenya Li, Hui Tian, and Shimin Cai.
\newblock Hybrid influence of degree and h-index in the link prediction of
  complex networks.
\newblock {\em EPL (Europhysics Letters)}, 122(6):68003, 2018.

\bibitem{PSLDW:2021:EPL}
Liming Pan, Hui-Juan Shang, Peiyan Li, Haixing Dai, Wei Wang, and Lixin Tian.
\newblock Predicting hyperlinks via hypernetwork loop structure.
\newblock {\em EPL (Europhysics Letters)}, 2021.

\end{thebibliography}
	\end{document}